\documentclass[conference]{IEEEtran}

\usepackage{cite}

\usepackage{graphicx}
\usepackage{xcolor}
\usepackage{tikz}
\usetikzlibrary{arrows,positioning,shapes.geometric}
\usepackage{pgfplots}
\usepackage{multirow}
\usepackage{upgreek}
\usepackage{amssymb}
\usepackage{amsmath}
\usepackage{cases}
\usepackage{pifont}
\usepackage{bm}
\usepackage{amsthm}

\usepackage{tabularx}
\usepackage{booktabs}
\usepackage{filecontents}
\usepackage{subcaption}
\newcolumntype{Y}{>{\centering\arraybackslash}X}

\usepackage[noend,ruled]{algorithm2e}

\usepackage{array}

\newcommand{\fixme}[2]{\ifx&#2&{\leavevmode\color{red}#1}\else{\leavevmode\color{red}FIXME\{}#1{\leavevmode\color{red}\}}\footnote{{\leavevmode\color{red}#2}}\PackageWarning{Fixme}{#1: #2}\fi}

\title{Efficient Bit-Channel Reliability Computation for Multi-Mode Polar Code Encoders and Decoders}

\author{Carlo~Condo, Seyyed~Ali~Hashemi, Warren~J.~Gross}
\thanks{C.~Condo, S.~A.~Hashemi and W.~J.~Gross are with the Department of Electrical and Computer Engineering, McGill University, Montr\'eal, Qu\'ebec, Canada. e-mail: carlo.condo@mail.mcgill.ca, seyyed.hashemi@mail.mcgill.ca, warren.gross@mcgill.ca.}

\begin{document}

\maketitle

\begin{abstract}

Polar codes are a family of capacity-achieving error-correcting codes, and they have been selected as part of the next generation wireless communication standard. Each polar code bit-channel is assigned a reliability value, used to determine which bits transmit information and which parity. Relative reliabilities need to be known by both encoders and decoders: in case of multi-mode systems, where multiple code lengths and code rates are supported, the storage of relative reliabilities can lead to high implementation complexity. In this work, observe patterns among code reliabilities. We propose an approximate computation technique to easily represent the reliabilities of multiple codes, through a limited set of variables and update rules. The proposed method allows to tune the trade-off between reliability accuracy and implementation complexity. An approximate computation architecture for encoders and decoders is designed and implemented, showing $50.7\%$ less area occupation than storage-based solutions, with less than $0.05$~dB error correction performance degradation.

\end{abstract}

\section{Introduction} \label{sec:intro}

Polar codes are a class of error-correcting codes proposed in \cite{arikan}, that can achieve capacity with a low-complexity encoding and decoding. Their construction exploits the channel polarization effect: this means that some of the channels through which codeword bits are transmitted (called bit-channels) are more reliable than others. Information bits are transmitted through the most reliable bit-channels, while the least reliable are set to a fixed value (frozen bits): the relative order of reliabilities is dependent on the code length and on the signal-to-noise ratio (SNR) for which the code has been constructed.

The number and position of information bits in a polar codeword needs to be known by both the encoder and the decoder. For encoders, information bits need to be correctly interleaved with frozen bits before encoding, and frozen bits need to be re-set halfway through systematic encoding \cite{Sarkis_TCOMM16}. Decoders targeting any decoding algorithm need to be aware of the bit arrangement as well \cite{arikan,sarkis,tal_list,hashemi_SSCL,hashemi_FSSCL}. Hardware implementations of encoders and decoders usually consider the frozen-information bit pattern as an input, and thus do not evaluate its storage or calculation cost. Many implementations of encoders and decoders target a single or a limited number of combinations of code lengths and rates, and a single SNR \cite{sarkis,Giard_Unrolled,Giard_Unrolled_MM,yuan_multibit,xiong_symbol}: thus, it is possible to store the frozen-information bit pattern for each supported code.
However, practical applications demand the support of a possibly large number of code lengths and rates, and various SNRs. Multi-mode decoders, and thus encoders, need to grant an even higher degree of flexibility than what can currently be achieved \cite{hashemi_SSCL_TCASI}. Within this framework, the direct storage of the bit pattern for each supported case can lead to unbearable implementation costs.

A few recent works address the problem of easy construction of polar code relative reliabilities. Partial orders in the reliability of polar code bit-channels were noticed in \cite{Schurch_ISIT16}, while a theoretical framework based on $\beta$-expansion for fast polar code construction has been proposed in \cite{He_ARXIV17}. While these approaches greatly reduce the computation complexity of the relative reliabilities, the direct implementation cost is still very high. 

In this work, we propose an approximate method to compute the relative reliability of polar codes, that can be implemented with considerably lower complexity than the direct storage of all values, along with a flexible architecture that can be used in both encoders and decoders. The trade-off between implementation complexity and degree of approximation can be tuned according to the application constraints.  

The rest of the paper is structured as follows. Section \ref{sec:prel} briefly introduces polar codes. Section \ref{sec:rel} describes the observed patterns in reliabilities and proposes the approximate computation method. Section \ref{sec:sim} details a case study and evaluates the impact of various approximations on the error-correction performance, whereas Section \ref{sec:arch} details an architecture for the implementation of the proposed technique, and provides implementation results. Finally, Section \ref{sec:conc} draws the conclusions.

\section{Polar Codes} \label{sec:prel}


A polar code $\mathcal{P}(N,K)$ is a linear block code of length $N=2^n$ and rate $K/N$. It is constructed as the concatenation of two polar codes of length $N/2$, and can be expressed as the matrix multiplication
\begin{equation}\label{eq:enc}
\mathbf{x} = \mathbf{u} \mathbf{G}^{\otimes n}
\end{equation}
where $\mathbf{u} = \{u_0,u_1,\ldots,u_{N-1}\}$ is the input vector, $\mathbf{x} = \{x_0,x_1,\ldots,x_{N-1}\}$ is the codeword, and the generator matrix $\mathbf{G}^{\otimes n}$ is the $n$-th Kronecker product of the polarizing matrix $\mathbf{G}=\bigl[\begin{smallmatrix} 1&0\\ 1&1 \end{smallmatrix} \bigr]$. The polar code structure allows to sort the $N$-bit input vector $\mathbf{u}$ according to the reliability of the bit-channels. The reliabilities associated with the bit-channels can be determined either by using the Bhattacharyya parameter \cite{arikan}, or through the direct use of probability function \cite{tal}.
The $K$ information bits are assigned to the most reliable bit-channels of $\mathbf{u}$, while the remaining $N-K$ (frozen bits) are set to a predetermined value, usually $0$.
Codeword $\mathbf{x}$ is transmitted through the channel, and the decoder receives the Logarithmic Likelihood Ratio (LLR) vector $\mathbf{y} = \{y_0,y_1,\ldots,y_{N-1}\}$.

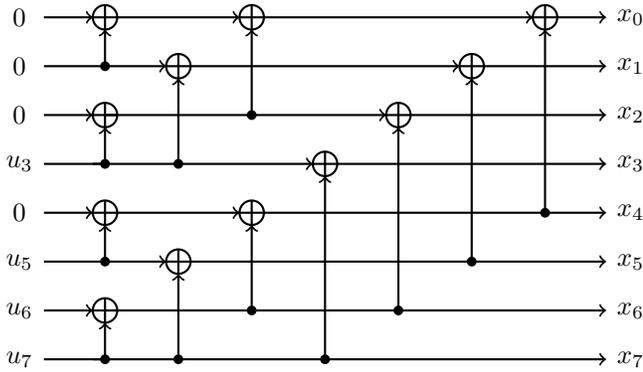
\begin{figure}
  \centering

\begin{tikzpicture}[scale=.65, thick]
  \node at (.5,0) {$0$};
  \node at (.5,-1) {$0$};
  \node at (.5,-2) {$0$};
  \node at (.5,-3) {$u_3$};
  \node at (.5,-4) {$0$};
  \node at (.5,-5) {$u_5$};
  \node at (.5,-6) {$u_6$};
  \node at (.5,-7) {$u_7$};

  \foreach \x in {-6,-4,-2,0}
  {
    \draw [->] (1,\x) -- (2,\x);
    \draw (1,\x-1) -- (2.25,\x-1);

    \draw (2.25,\x) circle [radius=.25];
    \draw (2,\x) -- (2.5,\x);
    \draw (2.25,\x-.25) -- (2.25,\x+.25);

    \draw [->] (2.25,\x-1) -- (2.25,\x-.25);

    \fill (2.25,\x-1) circle [radius=.1];
  }

  \foreach \x in {-4,0}
  {
    \draw [->] (2.5,\x) -- (5,\x);
    \draw [->] (2.25,\x-1) -- (3.5,\x-1);

    \draw (5.25,\x) circle [radius=.25];
    \draw (5,\x) -- (5.5,\x);
    \draw (5.25,\x-.25) -- (5.25,\x+.25);

    \draw (3.75,\x-1) circle [radius=.25];
    \draw (3.5,\x-1) -- (4,\x-1);
    \draw (3.75,\x-1-.25) -- (3.75,\x-1+.25);

    \draw [->] (2.25,\x-2) -- (5.25,\x-2) -- (5.25,\x-.25);
    \fill (5.25,\x-2) circle [radius=.1];
    \draw [->] (2,\x-3) -- (3.75,\x-3) -- (3.75,\x-1-.25);
    \fill (3.75,\x-3) circle [radius=.1];
  }

  \draw [->] (5.5,0) -- (11,0);
  \draw [->] (4,-1) -- (9.5,-1);
  \draw [->] (5.25,-2) -- (8,-2);
  \draw [->] (3.75,-3) -- (6.5,-3);

  \foreach \x in {-1,0}
  {
    \draw [->] (5.5+1.5*\x,\x-4) -- (11.25+1.5*\x,\x-4) -- (11.25+1.5*\x,\x-.25);
    \draw [->] (5.25+1.5*\x,\x-6) -- (11.25+1.5*\x-3,\x-6) -- (11.25+1.5*\x-3,\x-2-.25);
  }

  \foreach \x in {-3,...,0}
  {
    \draw (11.25+1.5*\x,\x) circle [radius=.25];
    \draw (11+1.5*\x,\x) -- (11.5+1.5*\x,\x);
    \draw (11.25+1.5*\x,\x-.25) -- (11.25+1.5*\x,\x+.25);

    \fill (11.25+1.5*\x,\x-4) circle [radius=.1];

    \draw [->] (11.5+1.5*\x,\x) -- (12.5,\x);
    \draw [->] (11.25+1.5*\x,\x-4) -- (12.5,\x-4);
  }

  \node at (13,0) {$x_0$};
  \node at (13,-1) {$x_1$};
  \node at (13,-2) {$x_2$};
  \node at (13,-3) {$x_3$};
  \node at (13,-4) {$x_4$};
  \node at (13,-5) {$x_5$};
  \node at (13,-6) {$x_6$};
  \node at (13,-7) {$x_7$};


\end{tikzpicture}

  \caption{Polar code encoding example for $\mathcal{P}(8,4)$ and $\{u_0,u_1,u_2,u_4\}\in \mathcal{F}$.}
  \label{figEnc}
\end{figure}

The encoding process in Equation~(\ref{eq:enc}) can be represented as in Fig.~\ref{figEnc}, that shows a polar code encoding example for $\mathcal{P}(8,4)$ where the frozen bits set $\mathcal{F}$ contains $\{u_0,u_1,u_2,u_4\}$.

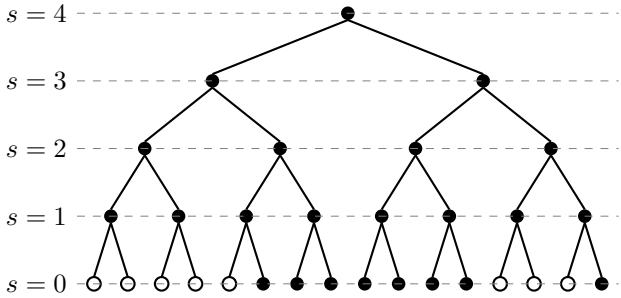
\begin{figure}
\centering
\begin{tikzpicture}[scale=1.8, thick]

\fill (0,0) circle [radius=.05];

\fill (-1,-.5) circle [radius=.05];
\fill (1,-.5) circle [radius=.05];

\fill (-1.5,-1) circle [radius=.05];
\fill  (-.5,-1) circle [radius=.05];
\fill  (.5,-1) circle [radius=.05];
\fill  (1.5,-1) circle [radius=.05];

\fill  (-1.75,-1.5) circle [radius=.05];
\fill  (-1.25,-1.5) circle [radius=.05];
\fill (-.75,-1.5) circle [radius=.05];
\fill (-.25,-1.5) circle [radius=.05];
\fill  (.25,-1.5) circle [radius=.05];
\fill  (.75,-1.5) circle [radius=.05];
\fill  (1.25,-1.5) circle [radius=.05];
\fill (1.75,-1.5) circle [radius=.05];

\draw (-1.875,-2) circle [radius=.05];
\draw  (-1.625,-2) circle [radius=.05];
\draw  (-1.375,-2) circle [radius=.05];
\draw  (-1.125,-2) circle [radius=.05];
\draw  (-.875,-2) circle [radius=.05];
\fill  (-.625,-2) circle [radius=.05];
\fill  (-.375,-2) circle [radius=.05];
\fill  (-.125,-2) circle [radius=.05];
\fill  (.125,-2) circle [radius=.05];
\fill  (.375,-2) circle [radius=.05];
\fill  (.625,-2) circle [radius=.05];
\fill  (.875,-2) circle [radius=.05];
\draw  (1.125,-2) circle [radius=.05];
\draw  (1.375,-2) circle [radius=.05];
\draw  (1.625,-2) circle [radius=.05];
\fill  (1.875,-2) circle [radius=.05];

\draw (0,-.05) -- (-1,-.45);
\draw (0,-.05) -- (1,-.45);

\draw (-1,-.55) -- (-1.5,-.95);
\draw (-1,-.55) -- (-.5,-.95);
\draw (1,-.55) -- (.5,-.95);
\draw (1,-.55) -- (1.5,-.95);

\draw  (-1.5,-1.05) -- (-1.75,-1.45);
\draw  (-1.5,-1.05) -- (-1.25,-1.45);
\draw (-.5,-1.05) -- (-.75,-1.45);
\draw (-.5,-1.05) -- (-.25,-1.45);
\draw  (.5,-1.05) -- (.25,-1.45);
\draw  (.5,-1.05) -- (.75,-1.45);
\draw  (1.5,-1.05) -- (1.25,-1.45);
\draw  (1.5,-1.05) -- (1.75,-1.45);

\draw  (-1.75,-1.55) -- (-1.875,-1.95);
\draw  (-1.75,-1.55) -- (-1.625,-1.95);
\draw  (-1.25,-1.55) -- (-1.375,-1.95);
\draw  (-1.25,-1.55) -- (-1.125,-1.95);
\draw  (-.75,-1.55) -- (-.875,-1.95);
\draw  (-.75,-1.55) -- (-.625,-1.95);
\draw (-.25,-1.55) -- (-.375,-1.95);
\draw (-.25,-1.55) -- (-.125,-1.95);
\draw  (.25,-1.55) -- (.125,-1.95);
\draw (.25,-1.55) -- (.375,-1.95);
\draw (.75,-1.55) -- (.625,-1.95);
\draw  (.75,-1.55) -- (.875,-1.95);
\draw  (1.25,-1.55) -- (1.125,-1.95);
\draw  (1.25,-1.55) -- (1.375,-1.95);
\draw  (1.75,-1.55) -- (1.625,-1.95);
\draw (1.75,-1.55) -- (1.875,-1.95);

\draw [very thin,gray,dashed] (-2,0) -- (2,0);
\draw [very thin,gray,dashed] (-2,-.5) -- (2,-.5);
\draw [very thin,gray,dashed] (-2,-1) -- (2,-1);
\draw [very thin,gray,dashed] (-2,-1.5) -- (2,-1.5);
\draw [very thin,gray,dashed] (-2,-2) -- (2,-2);

\node at (-2.3,0) {$s=4$};
\node at (-2.3,-.5) {$s=3$};
\node at (-2.3,-1) {$s=2$};
\node at (-2.3,-1.5) {$s=1$};
\node at (-2.3,-2) {$s=0$};

\end{tikzpicture}
\caption{Binary tree example for $\mathcal{P}(16,8)$. White circles at $s=0$ are frozen bits, black circles at $s=0$ are information bits.}
\label{fig:tree}
\end{figure}

Polar codes have been defined in \cite{arikan} together with the successive cancellation (SC) decoder: SC-based decoding process can be represented as a full binary tree search, in which the tree is explored depth first, with priority to the left branches. Fig.~\ref{fig:tree} shows an example of SC decoding tree for $\mathcal{P}(16,8)$, where nodes at stage $s$ contain $2^s$ bits. White leaf nodes are frozen bits, while black leaf nodes are information bits.

\begin{figure}
\centering
\begin{tikzpicture}[scale=.5]

\draw [very thin,gray,dashed] (-2,0) -- (2,0);
\draw [very thin,gray,dashed] (-2,-2) -- (2,-2);
\draw [very thin,gray,dashed] (-2,-4) -- (2,-4);

\node at (-3,0) {$s+1$};
\node at (-3,-2) {$s$};
\node at (-3,-4) {$s-1$};

\fill (0,0) circle [radius=.25];
\fill  (0,-2) circle [radius=.2];
\fill  (-1.5,-4) circle [radius=.15];
\fill  (1.5,-4) circle [radius=.15];

\draw [->,very thick] (-.1,-.4) -- (-.1,-1.7) node [left,midway,rotate=0] {$\alpha$};
\draw [->,very thick] (.1,-1.7) -- (.1,-.4) node [right,midway,rotate=0] {$\beta$};

\draw [->,very thick] (-.25,-2.2) -- (-1.45,-3.75) node [left,midway,rotate=0] {$\alpha^{\text{l}}$};
\draw [->,very thick] (-1.3,-3.85) -- (-.1,-2.3) node [right,near start,rotate=0] {$\beta^{\text{l}}$};
\draw [<-,very thick] (.25,-2.2) -- (1.45,-3.75) node [right,midway,rotate=0] {$\beta^{\text{r}}$};
\draw [<-,very thick] (1.3,-3.85) -- (.1,-2.3) node [left,near start,rotate=0] {$\alpha^{\text{r}}$};

\end{tikzpicture}
\caption{Message passing in tree graph representation of SC decoding.}
\label{fig:MessagePassing}
\end{figure}
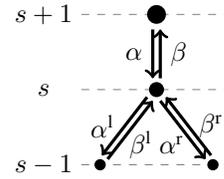

The message passing criteria among tree nodes is detailed in Fig.~\ref{fig:MessagePassing}. LLR values $\alpha$ are sent from parents to children, that in return send back the hard bit estimates $\beta$. 
Left branch messages $\alpha^\text{l}$ and right branch messages $\alpha^\text{r}$ can be computed in a hardware-friendly way \cite{leroux} as
\begin{align}
\alpha^{\text{l}}_i = & \text{sgn}(\alpha_i)\text{sgn}(\alpha_{i+2^{s-1}})\min(|\alpha_i|,|\alpha_{i+2^{s-1}}|) \\
\alpha^{\text{r}}_i =& \alpha_{i+2^{s-1}} + (1-2\beta^\text{l}_i)\alpha_i \text{,}
\label{eq2}
\end{align}
while $\beta$ is computed as
\begin{equation}
\beta_i =
  \begin{cases}
    \beta^\text{l}_i\oplus \beta^\text{r}_i, & \text{if} \quad i < 2^{s-1}\\
    \beta^\text{r}_{i-2^{s-1}}, & \text{otherwise},
  \end{cases}
  \label{eq3}
\end{equation}
where $\oplus$ is the bitwise XOR. Due to data dependencies, SC computations need to follow a particular schedule. Every node receives $\alpha$ first, then computes $\alpha^\text{l}$, receives $\beta^\text{l}$, computes $\alpha^\text{r}$, receives $\beta^\text{r}$, and finally sends back $\beta$.
At leaf nodes, $\beta_i$ is set as the estimated bit $\hat{u}_i$:
\begin{equation}
\hat{u}_i =
  \begin{cases}
    0 \text{,} & \text{if } i \in \mathcal{F} \text{ or } \alpha_{i}\geq 0\text{,}\\
    1 \text{,} & \text{otherwise.}
  \end{cases} \label{eq6}
\end{equation}

SC decoding suffers from mediocre error correction performance when applied to moderate and short code lengths. The SC-list (SCL) algorithm described in \cite{tal_list} improves the error correction performance by storing a set of $L$ codeword candidates, that gets updated after every bit estimation.

\section{Approximate Reliability Computation} \label{sec:rel}

\begin{figure*}
   \centering
      \includegraphics[scale=0.9]{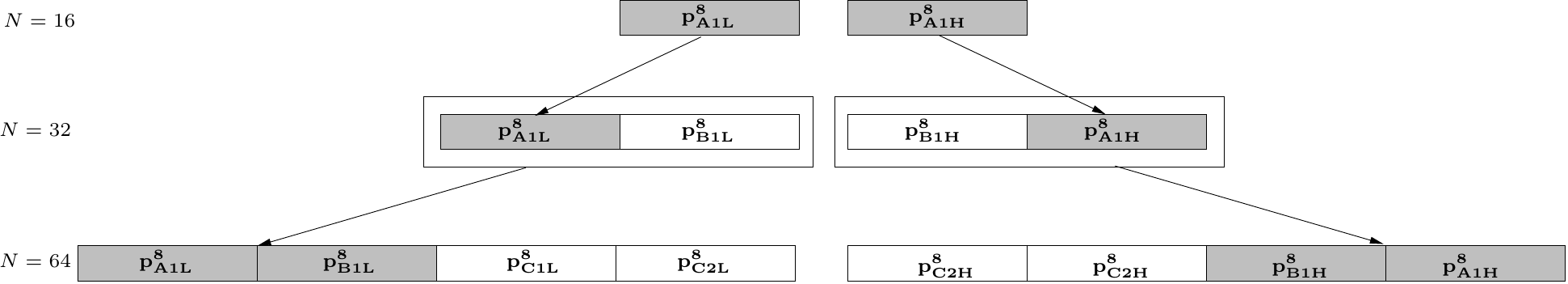}      
      \caption{Reliability bytes reusage scheme through different code lengths.}
      \label{fig:struct}
\end{figure*}

Given a polar code $\mathcal{P}(N,K)$, let us define the reliability vector $\mathbf{p}$ as the N-length sequence of elements $p_i$, where $0\le i<N$. Each element $p_i$ represents the reliability of bit-channel $i$, where $p_i=N-1$ is the least reliable bit and $p_i=0$ is the most reliable bit. Index $i=0$ refers to the leftmost bit on the decoding tree, while $i=N-1$ refers to the rightmost.

It is possible to identify regular patterns in the reliability vectors of polar codes, both within the same $\mathbf{p}$ and among vectors constructed for polar codes with different lengths. These patterns allow to efficiently describe $\mathbf{p}$ through variables, update rules and scaling.

In this section, we describe some of the patterns that we have identified and used to derive a hardware-efficient approximate reliability computation technique. We focused on codes constructed for the AWGN channel with the method used in \cite{tal}, targeting an SNR of around 6~dB. However, the proposed method can be easily extended to the reliabilities of codes constructed for other SNRs. 

We define here some variables that are going to be useful to explain the proposed method.
Let us divide the reliability vector $\mathbf{p}$ in two halves, $\mathbf{p_L}$ and $\mathbf{p_H}$, where $\mathbf{p_L}$ contains all $p_i$ for $0\le i <N/2$, and $\mathbf{p_H}$ the other ones. We call a reliability byte $\mathbf{p^8_L}$ a series of eight $p_i$ belonging to $\mathbf{p_L}$, and $\mathbf{p^8_H}$ one belonging to $\mathbf{p_H}$. To identify different reliability bytes, we assign an additional subscript to $\mathbf{p^8_L}$ and $\mathbf{p^8_H}$, as in $\mathbf{p^8_{B1L}}$ and $\mathbf{p^8_{E3H}}$. 

\subsection{Intra-code reliability patterns}\label{subsec:intra}

Observing $\mathbf{p_L}$ from $i=0$, it is possible to see how, generally, $p_i$ tends to decrease (i.e. becomes more reliable) as $i$ increases: this is because the first bits tend to be the least reliable of the code. In the same way, in $\mathbf{p_H}$, starting at $i=N-1$, $p_i$ tends to increase as $i$ decreases, since the last bits are usually the most reliable.

Both $\mathbf{p_L}$ and $\mathbf{p_H}$ can be expressed as a series of variables associated to update rules. As an example, let us consider $\mathbf{p}$ for $N=8$:
\begin{equation*} 
\mathbf{p}=\{7,6,5,3,4,2,1,0\} 
\end{equation*}
where $\mathbf{p_L}=\{7,6,5,3\}$ and $\mathbf{p_H}=\{4,2,1,0\}$. We can write these vectors as, for example:
\begin{equation*} 
\mathbf{p_L}=\{N-1, ~N-2, ~N-3,~ \text{ENDL}\} ,
\end{equation*}
\begin{equation*} 
\mathbf{p_H}=\{\text{ENDH}, ~H+2, ~H+1, ~H\}. 
\end{equation*}

As the code length increases, the regularity of the reliabilities decreases, and either a higher number of variables are needed to represent $\mathbf{p}$ exactly, or more irregular update patterns need to be used. For example, for $N=16$:
\begin{equation*} 
\mathbf{p}=\{15,  14,  13,  10 ,12 , 9 ,  8  , 4, 11,  7 ,  6 ,  3 ,  5, 2  ,1  , 0\}.
\end{equation*}
A possible representation of $\mathbf{p_L}$ and $\mathbf{p_H}$ can be:
\begin{equation*} 
\mathbf{p_L}=\{N-1, ~N-2, ~N-3, ~Z, ~N-4, ~Z-1, ~Z-2, ~ \text{ENDL}\} ,
\end{equation*}
\begin{equation*} 
\mathbf{p_H}=\{\text{ENDH}, ~I+2, ~I+1, ~H+3, ~I, ~H+2, ~H+1, ~H\}. 
\end{equation*}

\begin{table}[t!]
	\centering
	\caption{Variable sequence in $\mathbf{p}$ for $N=32$.} \label{tab:var32}
		\setlength{\extrarowheight}{1.7pt}
	\begin{tabular}{cc}
	\hline
	  Reliability byte & Composition sequence \\
	  \hline
	  \hline
	  $\mathbf{p^8_{A1L}}$ & $NNNZNZZY$ \\
	  $\mathbf{p^8_{B1L}}$ & $NZZYXYY ~\text{ENDL}$ \\
	  $\mathbf{p^8_{B1H}}$ & $\text{ENDH}~LLILIIH$ \\
	  $\mathbf{p^8_{A1H}}$ & $LIIHIHHH$ \\
	  	  \hline
	  \end{tabular}
\end{table}

With larger code lengths, we can derive an approximate $\mathbf{p}$ by limiting the number of variables used, and assigning a single, regular update pattern to each variable. As an example, Table \ref{tab:var32} reports the variable allocation within $\mathbf{p_L}=\{\mathbf{p^8_{A1L}},~\mathbf{p^8_{B1L}}\}$ and $\mathbf{p_H}=\{\mathbf{p^8_{B1H}}, ~\mathbf{p^8_{A1H}}\}$. An initialization value and a single update rule are selected for each variable: every time that a variable is encountered, the previous value is substituted with the updated one. For example, variable $Z$ is initialized as $25$, and its update rule is $-1$. Thus, $p_{3}=25$, $p_{5}=24$, $p_{6}=23$, $p_{9}=22$ and $p_{10}=21$. 

Moreover, the variable sequence within different reliability bytes of the same code can repeat itself, like in case of the second and third byte of $\mathbf{p_H}$ for $N\ge64$.



\subsection{Inter-code reliability patterns}\label{subsec:inter}

Having defined the code reliability as in Section \ref{subsec:intra}, it is possible to observe the evolution of $\mathbf{p}$ from a smaller code to larger codes.
A first observation can be made towards the reuse of $\mathbf{p_L}$ and $\mathbf{p_H}$ of a length-$N$ code as part of $\mathbf{p_L}$ and $\mathbf{p_H}$ of a length-$2N$ code. Fig. \ref{fig:struct} shows that the variable sequence found in $\mathbf{p_L}$ for length $N$, can approximate the first $N/2$ positions of $\mathbf{p_L}$ for length $2N$. A mirrored observation can be made for $\mathbf{p_H}$. Different initialization values and different update rules will be necessary when the code length is changed, but the same variable sequence can be used.
The precision with which variable sequences of lower-length codes can approximate part of larger-length code sequences depends on the number of variables. A high number of variables will guarantee very good precision and will allow to reuse a particular variable sequence for much larger codes. On the other hand, a large number of variable results in a higher degree of implementation complexity.

The frequency of occurrence and positioning of a variable within $\mathbf{p}$ can often be computed exactly. For example, variables $N$ and $H$ in Table \ref{tab:var32} will be encountered every $\delta_i=2\delta_{i-1}+1$ variables. Moreover, the initialization values of many variables can be expressed in function of the code length. Variable $I$ can be initialized as $\log_2(N)+1$, while $\text{ENDL}=\log_2(N)$ and $L\approx6\log_2(N)-8$.

We thus exploit the observed intra- and inter-code reliability patterns to propose an efficient way to store code reliabilities in decoder implementations. A single variable sequence $\mathbf{p}$ is selected, targeting the maximum code length supported. Shorter code lengths can be derived by considering only some $\mathbf{p^8}$, as in Fig. \ref{fig:struct}. To each variable is assigned an initialization value and an update value for each supported code length. The complete $\mathbf{p}$ can be constructed sequentially, starting from $i=0$ for $\mathbf{p_L}$ and from $i=N-1$ for $\mathbf{p_H}$. The code structure derived for a certain code length can be extended to that targeting different SNR points by a limited number of ad-hoc $\mathbf{p^8}$ substitutions.
The selection of variables and update rules can be helped by theoretical construction approaches like \cite{He_ARXIV17}. The technique proposed in our work is orthogonal to the construction method.

\section{Simulations and performance}\label{sec:sim}

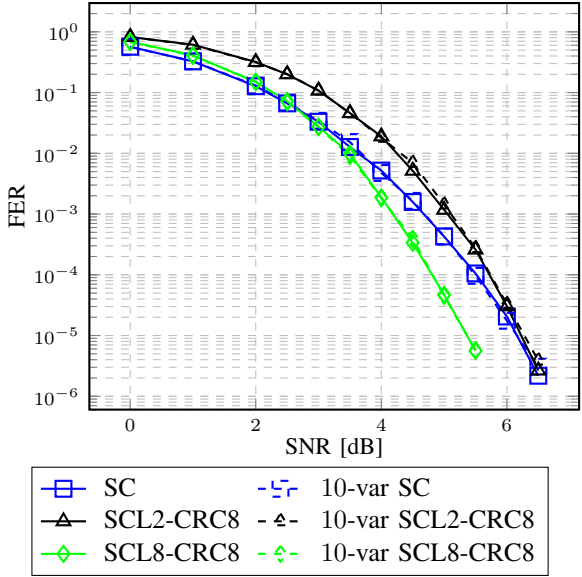
\begin{figure}
  \centering
  \begin{tikzpicture}
	{rectangle, magnification=2, connect spies}]
 \pgfplotsset{
   label style = {font=\fontsize{9pt}{7.2}\selectfont},
   tick label style = {font=\fontsize{7pt}{7.2}\selectfont}
 }

\begin{axis}
[
    ymode=log,
	scale = 0.95,
    xlabel={SNR [dB]}, xlabel style={yshift=0.8em},
    ylabel={FER}, ylabel style={yshift=-0.75em},%
    grid=both,
    ymajorgrids=true,
    xmajorgrids=true,
    grid style=dashed,
    thick,
    mark size=3,
    legend columns=2,
        legend style={
      cells={anchor=west},
      column sep= 1.5mm,
      font=\fontsize{10pt}{7.2}\selectfont,
    },
    legend to name=fer64legend
]

\addplot[
    color=blue,
    mark=square,
    thick,
    mark size=3,
]
table {
0 0.5678
1 0.3265
2 0.1272
2.5 0.0664
3 0.0331
3.5 0.0127
4 0.00519994
4.5 0.00158253
5 0.000427374
5.5 0.000105171
6 2.05716e-05
6.5 2.17835e-06
};
\addlegendentry{SC}

\addplot[
    color=blue,
    mark=square,
    thick,
    dashed,
    mark size=3,
]
table {
0 0.5642
1 0.3271
2 0.1305
2.5 0.0687
3 0.0343
3.5 0.0153
4 0.00475037
4.5 0.00164799
5 0.000428339
5.5 9.61982e-05
6 1.75134e-05
6.5 3.07416e-06 
};
\addlegendentry{$10$-var SC}

\addplot[
    color=black,
    mark=triangle,
    thick,
    mark size=3,
]
table {
0 0.8218
1 0.6143
2 0.3185
2.5 0.2001
3 0.1077
3.5 0.046
4 0.019
4.5 0.00516556
5 0.00117596
5.5 0.00026004
6 3.03614e-05
6.5 2.65239e-06
};
\addlegendentry{SCL2-CRC8}

\addplot[
    color=black,
    mark=triangle,
    thick,
    dashed,
    mark size=3,
]
table {
0 0.8239
1 0.6074
2 0.3153
2.5  0.2016
3 0.1081
3.5 0.0454
4 0.018
4.5 0.00694348
5 0.00143443
5.5 0.00027458
6 3.2994e-05
6.5 3.86223e-06
};
\addlegendentry{$10$-var SCL2-CRC8}

\addplot[
    color=green,
    mark=diamond,
    thick,
    mark size=3,
]
table {
0 0.6861
1 0.4111
2 0.1486
2.5 0.0725
3 0.0273
3.5 0.00984058
4 0.00188104
4.5 0.000336423
5 4.70114e-05
5.5 5.59848e-06
};
\addlegendentry{SCL8-CRC8}

\addplot[
    color=green,
    mark=diamond,
    thick,
    dashed,
    mark size=3,
]
table {
0 0.683
1 0.4026
2 0.1518
2.5 0.0693
3 0.027
3.5 0.0090009
4 0.00181478
4.5 0.000394215
5 4.47647e-05
5.5 5.77822e-06
};
\addlegendentry{$10$-var SCL8-CRC8}

\end{axis}
\end{tikzpicture}
  \\
  \ref{fer64legend}
  \caption{FER for $\mathcal{P}(64,32)$, for original reliabilities and approximated reliabilities with $10$ variables.}
  \label{fig:fer64}
\end{figure}

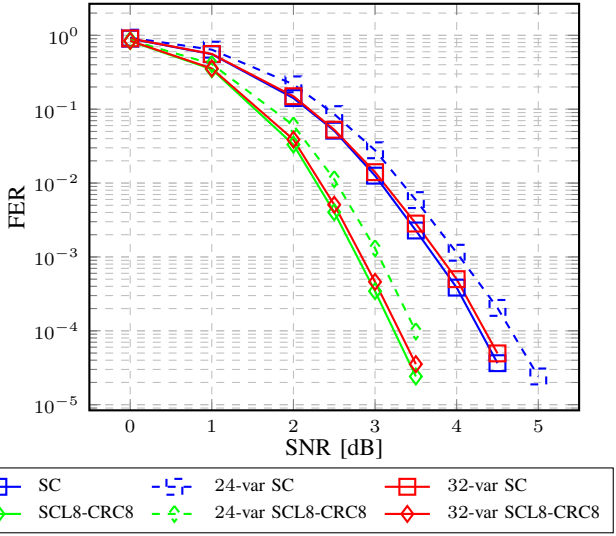
\begin{figure}
  \centering
  \begin{tikzpicture}
	{rectangle, magnification=2, connect spies}]
 \pgfplotsset{
   label style = {font=\fontsize{9pt}{7.2}\selectfont},
   tick label style = {font=\fontsize{7pt}{7.2}\selectfont}
 }

\begin{axis}
[
    ymode=log,
	scale = 0.95,
    xlabel={SNR [dB]}, xlabel style={yshift=0.8em},
    ylabel={FER}, ylabel style={yshift=-0.75em},%
    grid=both,
    ymajorgrids=true,
    xmajorgrids=true,
    grid style=dashed,
    thick,
    mark size=3,
    legend columns=3,
        legend style={
      cells={anchor=west},
      column sep= 1.5mm,
      font=\fontsize{7pt}{7.2}\selectfont,
    },
    legend to name=fer256legend
]

\addplot[
    color=blue,
    mark=square,
    thick,
    mark size=3,
]
table {
0 0.9021
1 0.5555
2 0.1414
2.5 0.0512
3 0.0126
3.5 0.0022804
4 0.00038442
4.5 3.64674e-05
};
\addlegendentry{SC}

\addplot[
    color=blue,
    mark=square,
    thick,
    dashed,
    mark size=3,
]
table {
0 0.9297
1 0.6396
2 0.2169
2.5 0.0858
3 0.0278
3.5 0.00588062
4 0.00113976
4.5 0.000204275
5 2.41015e-05
};
\addlegendentry{$24$-var SC}

\addplot[
    color=red,
    mark=square,
    thick,
    mark size=3,
]
table {
0 0.9086
1 0.5600
2 0.1498
2.5 0.0528
3 0.01405
3.5 0.0028313
4 0.0005001
4.5 4.94589e-05
};
\addlegendentry{$32$-var SC}

\addplot[
    color=green,
    mark=diamond,
    thick,
    mark size=3,
]
table {
0 0.8368
1 0.3475
2 0.0332
2.5 0.00402852
3 0.000344073
3.5 2.41025e-05
};
\addlegendentry{SCL8-CRC8}

\addplot[
    color=green,
    mark=diamond,
    thick,
    dashed,
    mark size=3,
]
table {
0 0.8625
1 0.409
2 0.0605
2.5 0.0113
3 0.00131206
3.5 9.844e-05
};
\addlegendentry{$24$-var SCL8-CRC8}

\addplot[
    color=red,
    mark=diamond,
    thick,
    mark size=3,
]
table {
0 0.8429
1 0.3561
2 0.0388
2.5 0.0050944
3 0.00045981
3.5 3.53995e-05
};
\addlegendentry{$32$-var SCL8-CRC8}

\end{axis}
\end{tikzpicture}
  \\
  \ref{fer256legend}
  \caption{FER for $\mathcal{P}(256,128)$, for original reliabilities and approximated reliabilities with $24$ and $32$ variables.}
  \label{fig:fer256}
\end{figure}

As a proof of concept, we provide the full construction method of the approximate reliability for codes of length 8 to 256. Table \ref{tab:struct} lists the variable sequences for the reliability bytes needed to construct codes of with maximum code length of 256. The boldfaced variables in the low (high) half are substituted with $\text{ENDL}$ ($\text{ENDH}$) when they represent $\mathbf{p_{N/2-1}}$ ($\mathbf{p_{N/2}}$). Initialization and update values for all considered code lengths are listed in Table \ref{tab:variables}. The choice of variable placement and their total number is one of the possible schemes: a larger number of variables will lead to more precise representation of $\mathbf{p}$, but will lead to higher storage requirements.

\begin{table}[t!]
	\centering
	\caption{Variable sequence in $\mathbf{p}$ for $8\le N \le 256$.}
	\label{tab:struct}
		\setlength{\extrarowheight}{1.7pt}
	\begin{tabular}{cc}
	\hline
	  Reliability byte & Composition sequence \\
	  \hline
	  \hline
	  $\mathbf{p^8_{A1L}}$ &  $NNN\mathbf{Z}NZZ\mathbf{Y}$ \\
	  $\mathbf{p^8_{B1L}}$ &  $NZZYXYY\mathbf{W}$ \\
	  $\mathbf{p^8_{C1L}}$ &  $NZXYXYYW$ \\
	  $\mathbf{p^8_{C2L}}$ &  $XYYWYWW\mathbf{V}$ \\
	  $\mathbf{p^8_{D1L}}$ &  $NXXY_2XY_2Y_2W$  \\
	  $\mathbf{p^8_{D2L}}$ &  $XY_2Y_2WY_2WWV$ \\ 
	  $\mathbf{p^8_{D3L}}$ &  $XY_2Y_2WY_2WWV$ \\
	  $\mathbf{p^8_{D4L}}$ &  $Y_2WWVWVV\mathbf{U}$ \\
	  $\mathbf{p^8_{E1L}}$ &  $NY_3Y_3Y_3Y_3TTW_2$	\\
	  $\mathbf{p^8_{E2L}}$ &  $Y_3TTW_2TW_2W_2V_2$	\\
	  $\mathbf{p^8_{E3L}}$ &  $Y_3TTW_2TSSV_2$	\\
	  $\mathbf{p^8_{E4L}}$ &  $TSSV_2SV_2V_2U$	\\
	  $\mathbf{p^8_{E5L}}$ &  $Y_3TTSTSSV_2$	\\
	  $\mathbf{p^8_{E6L}}$ &  $TSSV_2SV_2V_2U$	\\ 
	  $\mathbf{p^8_{E7L}}$ &  $TSSV_2SV_2V_2U$	\\ 
	  $\mathbf{p^8_{E8L}}$ &  $SV_2V_2UV_2UU~\text{ENDL}$	\\
	  \hline
	  $\mathbf{p^8_{E8H}}$ & $\text{ENDH}~QQQQQQM_2$	\\
	  $\mathbf{p^8_{E7H}}$ & $QO_2O_2M_2O_2M_2M_2L_2$	\\ 
	  $\mathbf{p^8_{E6H}}$ & $QO_2O_2O_2O_2M_2M_2L_2$	\\
	  $\mathbf{p^8_{E5H}}$ & $O_2M_2M_2L_2M_2L_2L_2I$	\\
	  $\mathbf{p^8_{E4H}}$ & $QO_2O_2M_2O_2M_2M_2L_2$	\\
	  $\mathbf{p^8_{E3H}}$ & $O_2M_2M_2L_2M_2L_2L_2I$	\\ 
	  $\mathbf{p^8_{E2H}}$ & $O_2M_2M_2L_2M_2L_2L_2I$	\\ 
	  $\mathbf{p^8_{E1H}}$ & $M_2L_2L_2IL_2IIH$	\\ 
	  $\mathbf{p^8_{D4H}}$ & $\mathbf{Q}OOMOMML$	\\
	  $\mathbf{p^8_{D3H}}$ & $OMMLMLLI$	\\ 
	  $\mathbf{p^8_{D2H}}$ & $OMMLMLLI$	 \\
	  $\mathbf{p^8_{D1H}}$ & $MLLILIIH$	\\
	  $\mathbf{p^8_{C2H}}$ & $\mathbf{O}MMLMLLI$	\\
	  $\mathbf{p^8_{C1H}}$ & $MLLILIIH$	\\ 
	  $\mathbf{p^8_{B1H}}$ & $\mathbf{M}LLILIIH$	\\
	  $\mathbf{p^8_{A1H}}$ & $\mathbf{L}IIH\mathbf{I}HHH$	\\
	  \hline
	  \end{tabular}
\end{table}

\begin{table}[t!]
	\centering
\caption{Variable initial values and update values for $\mathbf{p}$, $8\le N \le 256$.}
	\label{tab:variables}
		\setlength{\extrarowheight}{1.7pt}
	\begin{tabular}{cccc}
	\hline
	 \multirow{2}{*}{Variable} & \multirow{2}{*}{$N$} & Initial & \multirow{2}{*}{Update} \\
	 &&Value&\\
	 \hline
	 \hline
	 
	 $N$ & $8\rightarrow256$ & $N-1$ & $-1$ \\
	 \hline
	 $Z$ & $16\rightarrow256$ & $N-7$ & $-1$ \\
	 \hline
	 \multirow{4}{*}{$Y$} & $32$ & $15$ & $-1$ \\
	  & $64$ & $43$ & $-1.5$ \\
	 & $128$ & $108$ & $-2$ \\
	 & $256$ & $233$ & $-2.5$ \\
	 \hline
	 \multirow{2}{*}{$Y_2$}& $128$ & $95$ & $-2$ \\
	 & $256$ & $219$ & $-2.5$ \\
	 \hline
	$Y_3$ & $256$ & $232$ & $-4$ \\
	 \hline
	 \multirow{4}{*}{$X$} & $32$ & $19$ & -- \\
	  & $64$ & $50$ & $-1.5$ \\
	 & $128$ & $113$ & $-2$ \\
	 & $256$ & $241$ & $-2$ \\
	 \hline
	 \multirow{3}{*}{$W$} & $64$ & $21$ & $-1$ \\
	  & $128$ & $70$ & $-2$ \\
	 & $256$ & $204$ & $-5$ \\
	 \hline
	  $W_2$& $256$ & $174$& $-6$ \\
	 \hline
	 \multirow{2}{*}{$V$} & $128$ & $30$ & $-2$ \\
	 & $256$ & $140$ & $-10$ \\
	 \hline
	 $V_2$ & $256$ & $113$ & $-3$ \\
	 \hline
	 $U$& $256$& $47$ & $-3$ \\
	 \hline 
	 $T$& $256$& $204$ & $-3.5$ \\
	 \hline
	 $S$& $256$& $145$ & $-3$ \\
	 \hline
	 \hline
	 $H$ & $8\rightarrow256$ & $0$ & $+1$ \\
	 \hline
	 $I$ &  $16\rightarrow256$ & $\log_2(N)+1$ & $+1$ \\
	 \hline
	 \multirow{4}{*}{$L$} & $32$ & $16$ & $+1$ \\
	 & $64$ & $22$ & $+1$ \\
	 & $128$ & $28$ & $+1.5$ \\
	 & $256$ & $35$ & $+1.5$ \\
	 \hline
	 $L_2$& $256$ & $53$ & $+2$ \\
	 \hline
	 \multirow{3}{*}{$M$} & $64$ & $40$ & $+2$ \\
	 & $128$ & $61$ & $+2$ \\
	 & $256$ & $79$ & $+2$ \\
	 \hline
	 $M_2$ & $256$ & $95$ & $+3.5$ \\
	 \hline
	 \multirow{2}{*}{$O$} & $128$ & $91$ & $+3$ \\
	  & $256$ & $131$ & $+5$ \\
	  \hline
	  $O_2$ & $256$ & $143$ &  $+4$ \\
	  \hline
	  $Q$& $256$ & $192$ & $+3$ \\
	  \hline
	  \hline
	  $\text{ENDL}$ & $8\rightarrow256$ &  $I-1$ & -- \\
	  \hline
	  \multirow{6}{*}{$\text{ENDH}$} & $8$ &  $4$ &  --  \\
	  & $16$ &  $11$ &  --  \\
	  & $32$ &  $26$ & --  \\
	  & $64$ &  $56$ &  -- \\
	  & $128$ &  $116$ & --  \\
	  & $256$ &  $238$ & --  \\
	  
	  \hline
	\end{tabular}

\end{table}

The error-correction performance of the approximated reliability has been evaluated under both SC and SCL decoding. Fig. \ref{fig:fer64} shows the frame error rate (FER) for the $\mathcal{P}(64,32)$ polar code, both with the original reliabilities and the approximated ones according to Table \ref{tab:struct}-\ref{tab:variables}. The degradation in FER brought by the approximation is negligible for both SC and SCL ($L=2$ and $L=8$): thus, an approximation with 10 variables guarantees sufficient precision. Fig. \ref{fig:fer256} shows the same type of result for $\mathcal{P}(256,128)$: we can see that the proposed 24-variable approximation causes significant FER degradation, particularly severe with SC. Increasing the number of variables to $32$ allows a more precise approximation of $\mathbf{p}$: the red curves show a $\approx0.05$~dB FER degradation with respect to the ideal case. It is worth noting that the codes considered in Fig. \ref{fig:fer64}-\ref{fig:fer256} are rate 1/2, and thus more susceptible to the imprecisions of the approximation method. In fact, in general, the lowest and highest reliabilities are easy to represent with a regular structure, while the middle reliabilities are more complex. Thus, with high and low rate codes, the demarcation line between frozen and information bits will be closer to well-approximated reliabilities, and thus less likely to cause degradation.

\section{Reliability Computation Architecture} \label{sec:arch}

Reliability vectors need to be stored both at the encoder and the decoder side: multiple code lengths and rates can lead to high memory requirements. Frozen bit sequences require less memory to be stored, but a single reliability vector is sufficient for all code rates, while each frozen bit sequence identifies a single $\mathcal{P}(N,K)$. A frozen bit sequence is easily generated by comparing each reliability with a threshold value: if $p_i$ is higher than the threshold, then bit $i$ is among the least reliable bit-channels, and is used as a frozen bit.

\begin{figure}
   \centering
      \includegraphics[width=\columnwidth]{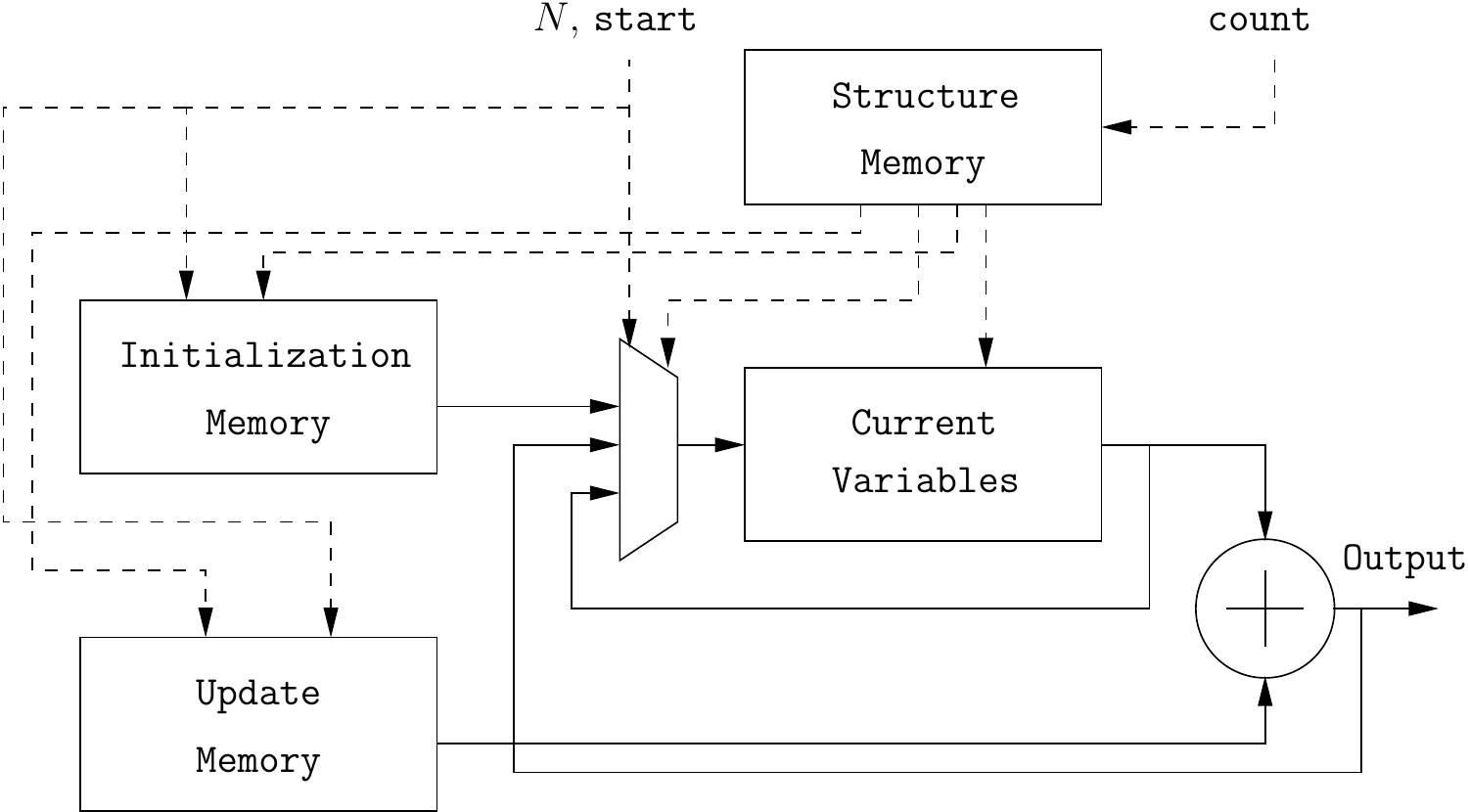}      
      \caption{Reliability computation architecture.}
      \label{fig:arch}
\end{figure}

The proposed approximated reliability computation method can be efficiently implemented in hardware.
Figure \ref{fig:arch} depicts the architecture at the encoder or decoder side: two of them are instantiated, one for $\mathbf{p_L}$ and one for $\mathbf{p_H}$. The \texttt{Structure memory} stores the variable sequence, as shown in Table \ref{tab:struct}. Each variable is represented with 5 bits, sufficient to represent up to $2^5$ variable types. The \texttt{Initialization memory} holds the 8-bit initialization values for all combinations of variables and code lengths in $\mathbf{p_L}$, and the \texttt{Update memory} the relative 5-bit update rules. In our case study, these two memories hold $38$ values for $\mathbf{p_L}$ and $30$ values for $\mathbf{p_H}$ in the $24$-variable case. Int the $32$-variable case, $43$ values are needed for $\mathbf{p_L}$ and $35$ for $\mathbf{p_H}$. 
A counter from $0$ to $\frac{N-1}{2}$ addresses the \texttt{Structure memory}: the output variable type, together with the code length, selects the correct variable and update values. The \texttt{current variables} memory holds the update value of all variable types used in the relative $\mathbf{p}$ half, appropriately initialized at the beginning of the computation. All memories can be reconfigured through external inputs.

The \texttt{Update memory} stores fixed point values with four integer bits and one decimal bit, while the \texttt{current variable} values are integers. Before the summation, the selected variable is left-shifted of one position, and the result is right-shifted back before the \texttt{current variables} memory is updated.

Along with the proposed architecture, we have designed a baseline architecture for comparative purposes. If no on-line construction technique is applied, the reliability vector for each supported code must be stored. Thus, the baseline considers a total $504$ 8-bit memory elements, along with the upload and reading logic. Both architectures output two reliabilities at each clock cycle.

Table \ref{tab:impl} reports synthesis results on 65~nm CMOS technology for the storage-based baseline architecture, along with the $24$-variable and the $32$-variable approximated architectures, supporting codes of length 8 to 256. The 24-var and 32-var results are relative to two instances of the architecture shown in Fig. \ref{fig:arch}, so that both $\mathbf{p_L}$ and $\mathbf{p_H}$ can be computed. With a target frequency of 500 MHz, the area occupation $A$ of the $24$-var architecture is $54.4\%$ less than that of the baseline. The multiplexing logic and adders account for a larger part of the area occupation than the baseline, reducing the percentage of memory elements $A_{mem}$. The $32$-var architecture extra memory requirements and logic lead to an area reduction $A_{red}$ of $50.7\%$. When the target frequency is increased to 1~GHz, $A_{red}$ decreases for both approximated architectures. This is due to the longer critical path that the reliability computation entails, that leads to higher synthesis effort and logic duplication.

\begin{table}[t!]
	\centering
	\caption{Implementation results for reliability storage and computation in CMOS $65$~nm technology.}
	\label{tab:impl}
	\renewcommand{\arraystretch}{1.5}
  	\begin{tabular}{cccc}
	\hline
	 & Full storage & $24$-var & $32$-var \\
	\hline
	\hline
	$f$ [MHz]& 500 & 500 & 500 \\
	$A$ [$\mu$m$^2$]  & $52817$ & $24103$	& $26039$	\\
	$A_{mem}$ [$\mu$m$^2$]  & $41249$ ($78.1\%$) 	& $17584$ ($73\%$)& $18941$ ($72.7\%$)	\\
	 $A_{red}$   & -- 	&	$-54.4\%$ & $-50.7\%$ \\
	 \hline
	$f$ [MHz]& 1000 & 1000 & 1000 \\
	 $A$ [$\mu$m$^2$]  & $52885$ & $33799$	& $35135$	\\
	$A_{mem}$ [$\mu$m$^2$]  & $41252$ ($78.0\%$) 	& $17871$ ($52.9\%$)& $19200$ ($54.6\%$)	\\
	 $A_{red}$   & -- 	&	$-36.1\%$ & $-33.6\%$ \\
	 
	\hline

	\end{tabular}
\end{table}

\section{Conclusion} \label{sec:conc}
In this work, we have proposed an approximate approach to polar code reliability computation that can efficiently be implemented in hardware encoders and decoders. Regular patterns within the reliability vector of a code and among those of different codes are observed, expressing the reliability vectors as a sequence of variables and update rules. Simulation show how the proposed method can be tuned to strike the desired trade-off between accuracy and ease of implementation. A low-complexity architecture is designed for various degrees of approximation, implemented and compared to a storage-based solution, showing $50.7\%$ less area occupation with $<0.05$~dB FER loss.

\bibliographystyle{IEEEtran}


\end{document}